\RequirePackage[final]{graphicx}
\documentclass[aps, superscriptaddress, prl, twocolumn, longbibliography]{revtex4-2}
\usepackage[utf8]{inputenc}
\usepackage{amsmath}
\usepackage{amssymb}
\usepackage{amsfonts}
\usepackage{mathtools}
\usepackage{fixme}
\usepackage{color}
\usepackage{braket}
\usepackage{dsfont}
\usepackage{hyperref}

\usepackage{tikz}
\usetikzlibrary{calc}

\usetikzlibrary{positioning,arrows}
\usetikzlibrary{decorations}
\usetikzlibrary{decorations.pathmorphing,positioning}
\usetikzlibrary{decorations.markings}

\tikzset
{
hole/.style         = { draw = black, postaction = { decorate }, decoration = { markings, mark = at position .55 with { \arrow[black]{ triangle 45} } } },
spinwave_11/.style      = { draw = cyan, postaction = { decorate }, decoration = { markings, mark = at position .5 with { \arrow[cyan]{ triangle 45} } } },
particle_small_arrow/.style         = { draw = black, postaction = { decorate }, decoration = { markings, mark = at position .5 with { \arrow[scale = 0.6, black]{ triangle 45} } } },
spinwave_12/.style      = { draw = cyan, postaction = { decorate }, decoration = { markings, mark = at position .25 with { \arrow[cyan, >=triangle 45]{<} }, mark = at position .75 with { \arrow[cyan, >=triangle 45]{>} } } },
spinwave_21/.style      = { draw = cyan, postaction = { decorate }, decoration = { markings, mark = at position .25 with { \arrow[cyan, >=triangle 45]{>} }, mark = at position .75 with { \arrow[cyan, >=triangle 45]{<} } } },
spinwave_total/.style           = { decorate, decoration = {snake, amplitude = 2pt, segment length = 7pt } }
}

\tikzset{
 ncbar angle/.initial=90,
 ncbar/.style={
 to path=(\tikztostart)
 -- ($(\tikztostart)!#1!\pgfkeysvalueof{/tikz/ncbar angle}:(\tikztotarget)$)
 -- ($(\tikztotarget)!($(\tikztostart)!#1!\pgfkeysvalueof{/tikz/ncbar angle}:(\tikztotarget)$)!\pgfkeysvalueof{/tikz/ncbar angle}:(\tikztostart)$)
 -- (\tikztotarget)
 },
 ncbar/.default=0.5cm,
}

\tikzset{square left brace/.style={ncbar=0.1cm}}
\tikzset{square right brace/.style={ncbar=-0.1cm}}

\definecolor{myred}{RGB}{214,26,70}
\definecolor{myreddark}{RGB}{76,8,38}
\definecolor{myblue}{RGB}{35,106,185}
\definecolor{mybluedark}{RGB}{19,56,99}
\definecolor{mybluebright}{RGB}{225,236,249}

\def\bi{{\bf i}}
\def\bj{{\bf j}}

\def\bsigma{{\pmb \sigma}}

\def\nn{\nonumber}

\def\FM{{ \rm FM }}
\def\Ham{{ \hat{H} }}

\begin{document}
\title{Dopant pairing in a disordered magnetic spin ladder}
\author{K.\ Knakkergaard Nielsen}
\affiliation{Max-Planck Institute for Quantum Optics, Hans-Kopfermann-Str. 1, D-85748 Garching, Germany}
\date{\today}

\begin{abstract}
I demonstrate a pairing mechanism of dopants in a magnetic lattice, emerging from the underlying \emph{high-temperature disorder} of the spins. The effect is demonstrated in a mixed-dimensional model, where dopants travel along a two-leg ladder, while the spins feature Ising interactions along the rungs and the legs of the ladder. The dynamics following the sudden immersion of two nearest neighbor dopants in an infinite temperature spin lattice shows that thermal spin disorder frustrates the relative motion of the dopants and enforces them to co-propagate. The predictions are shown to be realistically testable in quantum simulation experiments.
\end{abstract}

\maketitle

Understanding pairing of dopants in magnetic lattices is an immense challenge in modern quantum many-body theory. The main goal is to gain a deeper understanding of strongly correlated systems, arising e.g. in the notoriously difficult high-temperature superconductors \cite{highTc,Trugman1988,Dagotto1994}. While solid state experiments continue to pursue such inquiries after decades of research \cite{OMahony2022}, a highly interesting and promising pathway has appeared in quantum simulation experiments with ultracold atoms in optical lattices \cite{Gross2017, Ji2021, Wang2021, Prichard2024, Koepsell2019, Koepsell2021, Lebrat2024, Hirthe2023}. Single-site resolution capabilities \cite{Bakr2009,Sherson2010} makes it possible to measure low-temperature spatial correlations \cite{Koepsell2019,Koepsell2021,Prichard2024,Lebrat2024}, as well as track them over time by repeated realizations of the same initial state. In particular, single hole dynamics consistent with magnetic polaron formation \cite{Nielsen2022_2} in quantum antiferromagnets has been measured \cite{Ji2021}. Moreover, in such experiments one can tinker with the simulated models, otherwise fixed in the solid state. This has enabled the observation of dopant pairing in a mixed-dimensional magnetic lattice \cite{Hirthe2023}, also seen to be a plausible explanation \cite{Schlomer2024} for high-temperature superconductivity in a specific nickelate compound \cite{Sun2023}. More broadly speaking, the appearance of pairing is conventionally tied to the distortion of some underlying order, be it the underlying antiferromagnetic order in the above example \cite{Bohrdt2022, Nielsen2023_1}, or how, in conventional superconductors, electrons distort the underlying crystal lattice, inducing an attraction between them \cite{Frohlich1950,Bardeen1957_2,BruusFlensberg}. In materials supporting high-temperature superconductivity, there may be an important counterexample to this generic scenario, in which preformed Cooper pairs \cite{Geshkenbein1997,Randeria2001} appearing above the superconducting phase transition might explain \cite{Li2010,Seo2019} the pseudogap phase \cite{Stanescu2003}.
\begin{figure}[t!]
\begin{center}
\includegraphics[width=1.0\columnwidth]{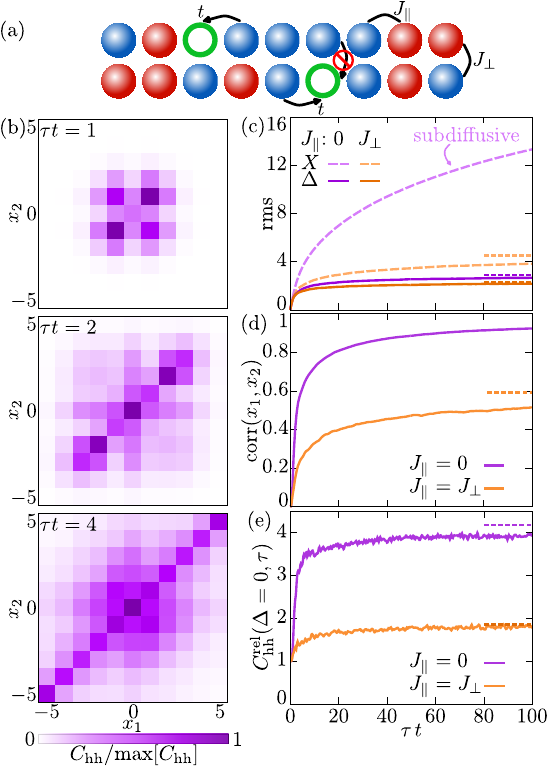}
\end{center}\vspace{-0.6cm}
\caption{{\bf Pairing dynamics.} (a) Setup: holes move in a magnetic (red: $\!\!\uparrow$, blue: $\!\!\downarrow$) two-leg ladder with Ising interactions $J_\perp,J_\parallel$ via nearest-neighbor hopping $t$ \emph{along} the ladder. (b) Hole-hole correlator $C_{hh}(x_1,x_2;\tau) = \braket{\hat{n}_h(x_1)\hat{n}_h(x_2)}_\tau$ for $J_\parallel = 0$ at indicated times normalized to its maximum after the holes are initialized on the same rung, $(l,x_1) = (1,0), (l,x_2) = (2,0)$. (c) Dynamics of the root-mean-square dynamics of the center-of-mass $X = (x_1 + x_2)/2$ and relative $\Delta = (x_1-x_2)/2$ coordinates, (d) correlation coefficient ${\rm corr}(x_1,x_2) = \braket{x_1x_2}_\tau/[\sigma(x_1)\sigma(x_2)]$, and (e) relative correlation function for $\Delta = 0$. Steady state values are shown in short dashed lines. In all: $J_\perp = 5t$.}
\label{fig.pairing_dynamics} 
\vspace{-0.25cm}
\end{figure} 

\begin{figure}[t!]
\begin{center}
\includegraphics[width=1.0\columnwidth]{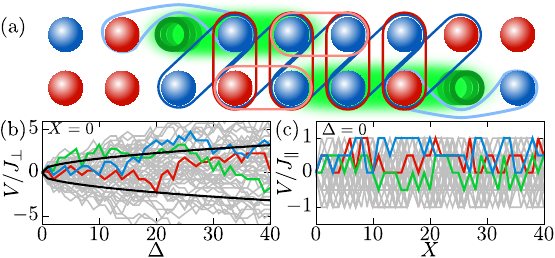}
\end{center}\vspace{-0.6cm}
\caption{{\bf Pairing mechanism.} (a) As the holes move (green shaded region), spin bonds in the initial configuration are broken across (dark blue) and within (light blue) the ladder, and new bonds are established (dark and light red, respectively), giving a random induced interaction between the holes. (b) Induced interaction for vanishing center-of-mass, $X = 0$, as a function of the relative coordinate $\Delta$ for $50$ realizations of the spin background (grey and colored lines), performing a random walk with standard deviation $\sim |J_\perp|\sqrt{|\Delta|}$ (black lines). (c) Induced interaction at $\Delta = 0$, as a function of $X$.}
\label{fig.potential_explanation} 
\vspace{-0.25cm}
\end{figure}

In this Letter, I demonstrate a pairing mechanism that \emph{does not} rely on being in nor in the vicinity of a phase transition to an ordered phase. In particular, the discovered mechanism remains at \emph{infinite} temperatures and is a direct consequence of the thermally disordered spin lattice. I analyze the effect in a simplistic setup of a mixed-dimensional model [Fig. \ref{fig.pairing_dynamics}(a)], where two holes are placed in two legs of a spin ladder. The spin degree of freedom is assumed to be at infinite temperature and completely disordered. Conventional wisdom, therefore, suggests that the holes should not be able to pair. Nevertheless, when the holes are initialized on the same rung in the infinite temperature spin lattice, the strong thermal disorder of the system frustrates the relative motion of the dopants. Indeed, tracking the hole-hole correlator in time [Fig. \ref{fig.pairing_dynamics}(b)], shows a significant correlation along the diagonal $x_1 = x_2$, where the holes are on the same rung of the ladder. One may equivalently observe that the inter-hole distance is always smaller than the traversed distance of their center-of-mass [Fig. \ref{fig.pairing_dynamics}(c)]. This results in a strictly \emph{positive} correlation between the holes' positions, signaling an effective attraction between them [Fig. \ref{fig.pairing_dynamics}(d)]. Finally, we observe that the probability of finding the holes on the same rung is easily increased by a factor of $4$, as compared to what one expects from their marginal distributions. See the relative hole-hole correlator in Fig. \ref{fig.pairing_dynamics}(e). This phenomenon arises due to an effective induced interaction that has a disordered character, encouraging the holes to co-propagate, and is closely tied to the recent discovery of thermally induced localization in such systems \cite{Nielsen2023_3,Nielsen2024_1}. In the extreme limit of $J_\parallel = 0$, the holes even dislodge from their original positions and spread \emph{subdiffusively as a pair}. I emphasize that the discovered effect would appear in a wide range of systems, though the robustness of the pairing in more generic setups is presently unclear.
\paragraph{Model.-} I consider spins in a two-leg ladder with Ising interactions and hopping \emph{along} the ladder,
\begin{align}
\Ham = \sum_{\bi,\bj} \!\frac{J_{\bi-\bj}}{2} \hat{S}^{(z)}_\bi\hat{S}^{(z)}_\bj - t \!\sum_{\braket{\bi,\bj}_\parallel,\sigma} \!\!\!\left[\tilde{c}^{\dagger}_{\sigma\bi}\tilde{c}_{\sigma\bj} + \tilde{c}^{\dagger}_{\sigma\bj}\tilde{c}_{\sigma\bi}\right],
\label{eq.H}
\end{align}
i.e. a mixed-dimensional $t$-$J_z$ model \cite{Grusdt2020}. I focus on the case of nearest-neighbor couplings along, $J_\parallel = J_{\pm {\bf e}_x}$, and across, $J_\perp = J_{\pm {\bf e}_y}$, the ladder, with $\hat{S}^{(z)}_\bi$ the $z$-projection of the spin operator $\hat{\bf S}_\bi$. The hopping, $\Ham_t$, is only allowed along the ladder and only to vacant sites, enforced by the constrained operators $\tilde{c}^\dagger_{\sigma\bi} = \hat{c}^\dagger_{\sigma\bi}[1 - \hat{n}_\bi]$, where $\hat{n}_\bi = \sum_{\sigma}\hat{n}_{\sigma\bi} = \sum_{\sigma}\hat{c}^\dagger_{\sigma\bi}\hat{c}_{\sigma\bi}$ is the local density operator. The effective spin degree of freedom generally describes an internal degree of freedom with the Ising-type coupling in $\Ham_J$, the first term in Eq. \eqref{eq.H}, and since there are no exchange processes of the dopants, both fermions and bosons can realize the model. To have an efficient description of the hole and spin degrees of freedom, I perform a Holstein-Primakoff transformation on top of the \emph{ferromagnetic} ground state $\ket{\FM} = \ket{\cdots \uparrow\uparrow\cdots}$, with all spins in the $\ket{\uparrow}$ state. The spin Hamiltonian
\begin{align}
\!\!\!\hat{H}_J \! = \!\sum_{\bi,\bj} \! \frac{J_{\bi-\bj}}{2} \Big[\frac{1}{2} \!-\! \hat{s}^\dagger_{\bi}\hat{s}_{\bi}\Big]\Big[\frac{1}{2} \!-\! \hat{s}^\dagger_{\bj}\hat{s}_{\bj}\Big] \Big[1 \!-\! \hat{h}_{\bi}^\dagger \hat{h}_{\bi}\Big] \Big[1 \!-\! \hat{h}_{\bj}^\dagger \hat{h}_{\bj}\Big],\!\!
\label{eq.H_J_holstein_primakoff}
\end{align}
hereby, describes the interaction of bosonic spin flip excitations $\hat{s}^\dagger_\bi$, corresponding to the creation of a spin-$\downarrow$ on site $\bi$, and holes created by the operator $\hat{h}^\dagger_\bi$, which inherit the statistics of the underlying particles, be it fermionic \emph{or} bosonic \cite{Nielsen2023_1}. Equivalently, the hopping Hamiltonian 
\begin{align}
\Ham_t = t \sum_{\braket{\bi,\bj}_\parallel} \! &\Big[ \hat{h}^\dagger_{\bj} F(\hat{h}_{\bi}, \hat{s}_{\bi}) F(\hat{h}_{\bj}, \hat{s}_{\bj}) \hat{h}_{\bi} \nn \\
&+ \hat{h}^\dagger_{\bj}\hat{s}^\dagger_\bi F(\hat{h}_{\bi}, \hat{s}_{\bi}) F(\hat{h}_{\bj}, \hat{s}_{\bj}) \hat{s}_\bj\hat{h}_{\bi} \Big] + {\rm H.c.}
\label{eq.H_t_holstein_primakoff}
\end{align}
describes two distinct ways in which holes may hop: (1) a hole hopping from site $\bi$ to $\bj$ in the absence of a spin flip on site $\bj$, and (2) hopping in the presence of a spin flip, whereby the hole and spin flip swap places. The constraint $F(\hat{h}, \hat{s}) = \sqrt{1 - \hat{s}^\dagger\hat{s} - \hat{h}^\dagger\hat{h}}$ ensures at most one spin per site. The initial state at time $\tau = 0$ of two holes on the same rung of the ladder in a spin background $\bsigma$ is then $
\ket{\Psi_\bsigma(\tau = 0)} = \prod_{l = 1}^{2} \hat{h}^\dagger_{(l,0)}\prod_{j\in S_{\bsigma}^l} \hat{s}^\dagger_{(l,j)} \ket{\FM}$, in which $S_{\bsigma}^l$ is the subset of spins in leg $l$, which are initially in the $\ket{\downarrow}$ state. Here, I write $\bi = (l,j)$, with leg index $l=1,2$ and position $j$ along the leg. Since spins only exchange places with the holes along the ladder, the state at any later stage is
\begin{align}
\ket{\Psi_\bsigma(\tau)} = &\sum_{x_1,x_2}\! a_\bsigma(x_1,x_2; \tau) \prod_{l = 1}^{2} \hat{h}^\dagger_{(l,x_l)} \nn \\
&\times\!\!\!\!\!\!\!\!\prod_{\substack{j \in S_{\bsigma}^l \\ j\in [{\rm sgn}(x_l),x_l]}} \!\!\!\!\!\!\!\hat{s}^\dagger_{(l,j-{\rm sgn}(x_l))} \!\!\!\!\!\prod_{\substack{j \in S_{\bsigma}^l \\ j\notin [{\rm sgn}(x_l),x_l]}} \!\!\!\!\!\hat{s}^\dagger_{(l,j)}\!\ket{\FM}.
\label{eq.Psi_sigma}
\end{align}
Here, ${\rm sgn}(x)$ is the sign function. Despite the rather formal description, the physics is simple and clear: if the hole in leg $l$ moves $|x_l|$ times to the right (left), it surpasses $|x_l|$ spins that all move one step to the left (right). The corresponding probability amplitudes $a_{\bsigma}(x_1,x_2; \tau)$ of finding the holes at positions $x_1,x_2$ at time $\tau$ obey
\begin{align}
i\partial_\tau a_\bsigma(x_1,x_2; \tau) &= V_\bsigma(x_1,x_2) a_{\bsigma}(x_1,x_2; \tau) \nn \\
&+ t\!\!\!\!\!\!\!\sum_{\substack{\delta_1,\delta_2 \\ |\delta_1| + |\delta_2| = 1}} \!\!\!\!\!\!\!a_{\bsigma}(x_1 + \delta_1,x_2 + \delta_2; \tau),
\label{eq.equations_of_motion}
\end{align}
derived from the Schr{\"o}dinger equation, $i\partial_\tau \ket{\Psi_\bsigma(\tau)} = \hat{H}\ket{\Psi_\bsigma(\tau)}$, with the initial condition that holes are on the same rung: $a_\bsigma(x_1 = 0,x_2 = 0; \tau = 0) = 1$. The sum extends over configurations reached in a single hop. The induced interaction $V_\bsigma(x_1,x_2)$ emerges, because the motion of the holes alter the spin bonds in the ladder [Fig. \ref{fig.potential_explanation}(a)]. For a random spin background, the induced interaction is also random [Fig. \ref{fig.potential_explanation}(b)], characterized in the relative and center-of-mass coordinates $\Delta = (x_1 - x_2) / 2$ and $X = (x_1 + x_2)/2$. 
I indeed show \cite{SM} that if all spin backgrounds are equally likely -- as in an infinite temperature spin state -- $V_\bsigma(x_1,x_2)$ has zero mean and a variance ${\rm Var}[V_{\bsigma}(x_1,x_2)] \propto J_\perp^2 |\Delta|$, linear in the relative distance $2|\Delta|$ between the holes, describing a classical random walk of $V$ in $|\Delta|$ [Fig. \ref{fig.potential_explanation}(b)]. I then focus on the evolution in specific spin backgrounds by resolving the dynamics of the density matrix in terms of these states. Starting from the initial density matrix
\begin{equation}
\hat{\rho}(0) = \!\sum_{\sigma_1,\sigma_2} \hat{c}_{(2,0),\sigma_{2}}\hat{c}_{(1,0),\sigma_1} \hat{\rho}_J \hat{c}_{(1,0),\sigma_1}^\dagger \hat{c}_{(2,0),\sigma_{2}}^\dagger,
\end{equation}
where $\hat{\rho}_J = e^{-\beta \Ham_J} / Z_J$ is the Gibbs state of the spins at inverse temperature $\beta = 1/(k_BT)$ and $\sigma_l = \uparrow,\downarrow$ designates the spin configuration at the origin of each leg, $l$, the ensuing dynamics $\hat{\rho}(\tau) = e^{-i\hat{H}\tau} \hat{\rho}(0) e^{+i\hat{H}\tau}$ -- provided that the system is assumed to be closed -- is the Boltzmann-weighted sum $\hat{\rho}(\tau) = Z_J^{-1}\sum_{\bsigma} e^{-\beta E_J(\bsigma)} \ket{\Psi_{\bsigma}(\tau)}\bra{\Psi_{\bsigma}(\tau)}$ \cite{Nielsen2023_3,Nielsen2024_1}, where $E_J(\bsigma)$ is the magnetic energy of the spin realization $\bsigma$ before the spins $\sigma_1, \sigma_2$ are removed. At infinite temperatures all pure state evolutions $\ket{\Psi_{\bsigma}(\tau)}$ are sampled equally, $e^{-\beta E_J(\bsigma)}/Z_J = 2^{-N}$, for system size $N$, and the ensuing dynamics is \emph{exactly equal} for antiferromagnetic ($J_\perp,J_\parallel > 0$) and ferromagnetic couplings ($J_\perp,J_\parallel < 0$). The calculation of the hole dynamics then follows the recipe: (1) generate a sample $S$ of $N_S = 400$ random spin configurations $\bsigma$ and calculate $V_{\bsigma}(x_1,x_2)$, (2) solve the equations of motion in Eq. \eqref{eq.equations_of_motion} using exact diagonalization for $N = 201\times 2$, and (3) compute the hole-hole correlator
\begin{equation}
C_{hh}(x_1,x_2; \tau) = \braket{\hat{n}_h(x_1)\hat{n}_h(x_2)}_\tau = \!\sum_{\bsigma\in S} \!\frac{|a_{\bsigma}(x_1,x_2; \tau)|^2}{N_S},
\end{equation}
as a function of time [Fig. \ref{fig.pairing_dynamics}(b)]. This describes the probability of simultaneously finding holes at positions $(1,x_1), (2,x_2)$, with $\hat{n}_h(x_l) = \hat{h}^\dagger_{(l,x_l)}\hat{h}_{(l,x_l)}$ the hole density operator. Figure \ref{fig.pairing_dynamics}(b) manifestly breaks the individual inversion symmetries $x_1\to-x_1$ \emph{or} $x_2 \to -x_2$ of uncorrelated holes, and shows a clear tendency for them to be on the same rung ($x_1 = x_2 \Leftrightarrow \Delta = 0$). Their motion is further analyzed by the root-mean-square of the center-of-mass and relative coordinates [Fig. \ref{fig.pairing_dynamics}(c)],
\begin{align}
A_{\rm rms}(\tau) = \left[\sum_{X,\Delta} A^2 C_{hh}(X+\Delta,X-\Delta; \tau)\right]^{1/2},
\end{align}
for $A = X, \Delta$. Since $X_{\rm rms} > \Delta_{\rm rms}$, the holes are closer together than they have propagated into the lattice. This is further characterized by the correlation coefficient
\begin{align}
{\rm corr}(x_1,x_2) = \frac{\braket{x_1x_2}_\tau}{\sigma(x_1)\sigma(x_2)} = \frac{X_{\rm rms}^2-\Delta_{\rm rms}^2}{X_{\rm rms}^2+\Delta_{\rm rms}^2},
\label{eq.normalized_covariance}
\end{align}
taking values in $[-1,+1]$. Positive values, as in Fig. \ref{fig.pairing_dynamics}(d), correspond to $X_{\rm rms} > \Delta_{\rm rms}$, and signify \emph{attraction}. Likewise, negative values correspond to effective repulsion, while ${\rm corr}(x_1,x_2) = 0$ corresponds to no linear correlations. To further substantiate the findings, I analyze the relative hole-hole correlator
\begin{equation}
C_{hh}^{\rm rel}(\Delta,\tau) = \frac{\sum_{X} C_{hh}(X+\Delta,X-\Delta; \tau)}{C_{hh}^{\rm disc}(\Delta, \tau)},
\label{eq.C_hh_rel}
\end{equation}
giving the probability of finding the holes at distance $2|\Delta|$ \emph{relative} to what is expected from their marginals, i.e. the disconnected correlator $C^{\rm disc}_{hh}(\Delta,\tau) = \sum_{X}\braket{\hat{n}_h(X+\Delta)}\braket{\hat{n}_h(X-\Delta)}$. This is shown for $\Delta = 0$ in Fig. \ref{fig.pairing_dynamics}(e), demonstrating a dynamically increasing probability of finding the holes on the same rung. 

\begin{figure}[t!]
\begin{center}
\includegraphics[width=1.0\columnwidth]{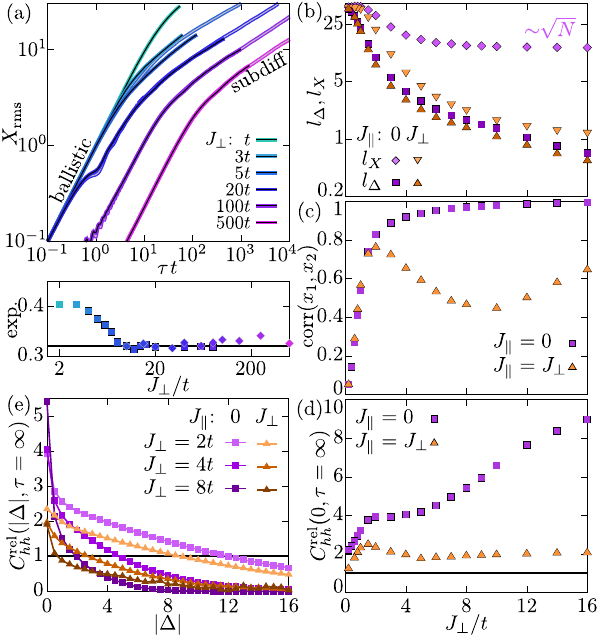}
\end{center}\vspace{-0.6cm}
\caption{{\bf Pair subdiffusion and steady state.} (a) For $J_\parallel = 0$, the hole pair spreads subdiffusively, $X_{\rm rms} \sim \tau^\nu$ at long times [top] with an exponent $\nu$ decreasing with $J_\perp/t$ [bottom]. Black- and gray-filled lines [top] give the dynamics for the full [Eq. \eqref{eq.H}] and effective models, with corresponding black- and gray-enscribed points for the fitted exponents [bottom]. (b) Steady state rms distances $l_X$ and $l_\Delta$ of the center-of-mass, $X$, and relative, $\Delta$, coordinates of the holes as a function of $J_\perp / t$ for specified $J_\parallel$. For $J_\parallel = 0$, $l_X$ saturates at $\sim\sqrt{N}$. (c) Correlation coefficient ${\rm corr}(x_1,x_2) = \braket{x_1x_2}/[\sigma(x_1)\sigma(x_2)]$, and (d) relative hole-hole correlator [Eq. \eqref{eq.C_hh_rel}] for holes on the same rung ($\Delta = 0$) vs $J_\perp / t$. (e) Relative hole-hole correlator vs $|\Delta|$ for indicated values of $J_\perp / t$ and $J_\parallel$.}
\label{fig.steady_state} 
\vspace{-0.25cm}
\end{figure}

For vanishing intra-leg spin coupling, $J_\parallel = 0$, the holes will continue to spread into the system. Indeed, by analyzing the center-of-mass dynamics [black-filled colored lines in Fig. \ref{fig.steady_state}(a)], I find that it transitions from an initial ballistic expansion with velocity $t$ to \emph{subdiffusive} behavior, $X_{\rm rms} \sim (\tau \cdot t)^\nu$. This persistent spreading comes from a vanishing potential for holes on the same rung, $V(x,x) = 0$. In fact, in the limit of $J_\perp \gg t$, I gain simplified quantitative insights by assuming the holes move at most one site away from each other. The resulting effective model allows me to push the numerics to larger systems and longer times, verifying the subdiffusive behavior, and is seen to perfectly match the full dynamics for large $J_\perp / t$ on all timescales [gray-filled colored lines in Fig. \ref{fig.steady_state}(a)]. The subdiffusive exponent [bottom of Fig. \ref{fig.steady_state}(a)] slowly decreases for increasing $J_\perp / t$, reaching a plateau at $\nu \simeq 0.32$ for $J_\perp / t \gg 1$.

Eventually, the system settles into a steady state, such that $X_{\rm rms} \to l_X, \Delta_{\rm rms} \to l_\Delta$ [Fig. \ref{fig.steady_state}(b)]. The associated asymptote of the correlation coefficient [Fig. \ref{fig.steady_state}(c)] shows attractive correlations over the entire investigated range. Moreover, the relative hole-hole correlator in Figs. \ref{fig.steady_state}(d) and \ref{fig.steady_state}(e) show that the probability of finding holes in close proximity can be up to an order of magnitude larger than expected from their marginals. For $J_\perp = J_\parallel \geq 3t$, there are minor finite-size corrections, whilst smaller spin couplings results in more spread out holes and significant corrections \cite{SM}. Indeed, for $J_\parallel = 0$, the subdiffusive behavior leads to a system-size dependent steady state, and I find $l_X \sim \sqrt{N}$. In turn, the correlation coefficient [Fig. \ref{fig.steady_state}(b)] behaves as $1 - {\rm corr}(x_1,x_2) \propto l_\Delta^2/N$. The relative hole-hole correlator is, however, found to be well-converged for any $J_\perp > 3t$ \cite{SM}. For $J_\perp \leq 3t$ and both values of $J_{\parallel}$, there are substantial finite-size effects that suppress the pairing signals, because it foremost restricts the center-of-mass motion. In this manner, pairing is demonstrating across the entire investigated regime and persists for $J_\perp, J_\parallel < t$. 

The origin of this dynamical pairing is \emph{exactly} the same as for the thermal localization phenomenon discovered recently for the motion of single holes \cite{Nielsen2023_3,Nielsen2024_1}. Indeed, because the induced interaction $V_\bsigma(x_1,x_2)$ inevitably fluctuates to large values in the relative distance $2|\Delta| = |x_1-x_2|$ [Fig. \ref{fig.potential_explanation}(b)], the holes, at the very least, back-reflect once they meet their classical turning points, corresponding to emergent Anderson localization \cite{Anderson1958} for \emph{strong disorder} \cite{Lagendijk2009}. For nonzero $J_\parallel$, the holes remain localized -- a finite $l_X$, because even when they are on the same rung, they will encounter a fluctuating potential taking on values in the interval $[-|J_\parallel|,+|J_\parallel]$ [Fig. \ref{fig.potential_explanation}(c)], leading to Anderson localization for \emph{weak} disorder \cite{Lagendijk2009}. 

\paragraph{Rydberg-dressed atoms.-} Quantum simulators based on Rydberg-dressed atoms in optical lattices natively realizes spin-specific density interactions \cite{Zeiher2016}
\begin{equation}
\Ham_J = \sum_{\bi \neq \bj} \frac{J_{\bi-\bj}}{2} \hat{n}_{\uparrow\bi}\hat{n}_{\uparrow\bj}
\label{eq.H_J_rydberg}
\end{equation}
of the internal atomic state $\ket{\uparrow}$, by optically dressing it with a higher-lying Rydberg state. Here, $J_{\bi-\bj}$ takes on a soft core shape \cite{Henkel2010,SM}. Such systems were recently demonstrated to enter the itinerant regime \cite{Weckesser2024}. Finally, a Feshbach resonance used to enhance the onsite interaction deep into the Mott-insulating phase ensures at most one spin per site, negligible superexchange, and hereby realizes the desired itinerant spin model governed by $\Ham_J$ in Eq. \eqref{eq.H_J_rydberg} and the constrained hopping $\Ham_t$ in Eq. \eqref{eq.H}. Precise control of the temperature in such systems is difficult, but the hole dynamics may still be investigated in the following manner \cite{Nielsen2023_3}. (1) initialize holes at $x_1 = x_2 = 0$ by applying a strong repulsive light field to those sites \cite{Ji2021}, (2) rotate all spins from the fully polarized spin sample in the non-interacting $\ket{\downarrow}$ state into the equal superposition state $\ket{\Psi_0} = \prod_{\bi \neq (0,0),(0,1)} \frac{1}{\sqrt{2}}\left[\hat{c}^\dagger_{\bi\uparrow} + \hat{c}^\dagger_{\bi\downarrow}\right]\ket{0}$, and (3) turn off the focused light fields, freeing the holes to move along the ladder \cite{Hirthe2023}. 

\begin{figure}[t!]
\begin{center}
\includegraphics[width=1.0\columnwidth]{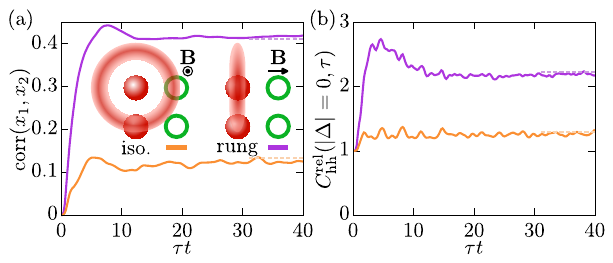}
\end{center}\vspace{-0.6cm}
\caption{{\bf Rydberg dressing.} (a) Dynamics of the correlation coefficient for isotropic (orange) and strong rung (purple) couplings. (b) Relative correlator for holes on the same rung, $\Delta = 0$. Dashed lines: steady state values. Nearest neighbor rung interaction: $J_{{\bf e}_y} = 8t$. Insets: $30P_{3/2}$ orbital orientation dictated by the magnetic fields, ${\bf B}$. System size: $N = 21\times 2$.}
\label{fig.rydberg_dynamics} 
\vspace{-0.25cm}
\end{figure}

This initially seems distinct from the infinite temperature scenario above. The hole-hole correlator dynamics, $C_{hh}(x_1,x_2; \tau) = 2^{-N}\sum_{\bsigma} |a_\bsigma(x_1,x_2;\tau)|^2$, is however unchanged, whereby the holes move as at infinite temperature. The polarized Ising form of the interaction is crucial for this exact equivalence, and one should note that \emph{spin} correlators would presumably differ from their infinite temperature counterpart. The computation now follows the same recipe as before. The resulting correlation coefficient dynamics [Eq. \eqref{eq.normalized_covariance}] and relative correlation function [Eq. \eqref{eq.C_hh_rel}] are given in Figs. \ref{fig.rydberg_dynamics}(a) and \ref{fig.rydberg_dynamics}(b), respectively. This is averaged over $N_S = 1000$ realizations and given in two instances [inset of Fig. \ref{fig.rydberg_dynamics}(a)]: (1) isotropic couplings akin to the $J_\parallel = J_\perp$ case, and (2) anisotropic couplings where the $30P_{3/2}$ orbital of the Rydberg state is rotated to lie along the rungs using an external magnetic field \cite{Zeiher2016}, suppressing coupling in other directions similar to low values of $J_\parallel$ \cite{SM}. Crucially, we observe that the appearing disorder-induced attraction between the holes can be probed for experimentally feasible system sizes \emph{and} timescales. Indeed, stroboscopic dressing \cite{Weckesser2024} makes $20$ tunneling events feasible, while longer timescales are currently inhibited by the intrinsic lifetime of the Rydberg state \cite{Priv_Comm_Weckesser}. 
\paragraph{Conclusions and outlook.-} I have demonstrated that thermal spin disorder can frustrate the relative motion of dopants and lead to dynamical pairing. The nature of the effect is very general: it comes down to a randomly fluctuating field emerging as the dopants separate. It is closely related to the thermally induced localization phenomenon recently discovered for single holes \cite{Nielsen2023_3,Nielsen2024_1}. By carefully analyzing its origin and behavior at infinite temperature, it is clear that the effect persists for all values of the spin coupling versus the hopping. For diminishing temperatures, ferromagnetic interactions should lead to a crossover to delocalized and uncorrelated holes. For antiferromagnetic interactions, a crossover from fluctuation induced pairing to \emph{confinement} of the holes emerges at low temperatures \cite{Nielsen2023_1}. As demonstrated, my predictions are feasibly testable using available techniques with ultracold atoms in optical lattices. Alternative platforms include Rydberg arrays \cite{Schauss2015,Labuhn2016,Bernien2017,Guardado-Sanchez2018,Lienhard2018} and polar molecules \cite{Gorshkov2011_2}. The generality of the discovered effect may be tested by adding trans-leg hopping, $t_\perp$, or spin superexchange $\propto \alpha J$. I expect pairing to be robust as long as the hopping along the ladder dynamically dominates the perturbations, $t \gg t_\perp, \alpha J$. Such computations go beyond the presently described methodology, but may be amenable to DMRG approaches \cite{White2015}. This will be addressed in future work. 

\begin{acknowledgments}
The author thanks Pascal Weckesser, Johannes Zeiher, Pavel Kos, and J. Ignacio Cirac for helpful input and discussions. Special thanks go to Pascal Weckesser for computing the soft core potentials. This work has been supported by the Carlsberg Foundation through a Carlsberg Internationalisation Fellowship, Grant No. CF21\_0410.
\end{acknowledgments} 

The data that support the findings of this article are openly available \cite{data_availability}.

\bibliography{ref_pairing_by_disorder}

\end{document}


\title{Supplemental Material \\ Pairing by disorder of dopants in magnetic spin ladder}

\maketitle

\beginsupplement
\tableofcontents

\section{Induced interaction of two holes} \label{sec.two_hole_interaction}
In this section, I write down the explicit induced interaction of two holes for a specific spin realization $\bsigma$. The role of this potential is that it directly enters the equations of motion for said spin realization,
\begin{align}
i\partial_\tau a_\bsigma(x_1,x_2; \tau) &= V_\bsigma(x_1,x_2) a_\bsigma(x_1,x_2; \tau) \nn \\ &+t \left[a_\bsigma(x_1-1,x_2; \tau) + a_\bsigma(x_1+1,x_2; \tau) + a_\bsigma(x_1,x_2 - 1; \tau) + a_\bsigma(x_1,x_2 + 1; \tau)\right],
\end{align}
as the effective local energy shift of the two holes at positions $(x_1,x_2)$. As we shall shortly see, there are two special features of this interaction: (1) it is not translationally invariant for general $\bsigma$, (2) it favors the holes to co-propagate. As for the single-hole case \cite{Nielsen2023_3}, I separate the potential into parts coming from the intra-leg and trans-leg spin couplings: $V_\bsigma = V_{\bsigma,\parallel} + V_{\bsigma,\perp}$. Here, $V_{\bsigma,\parallel} \propto J_\parallel$ and $V_{\bsigma,\perp} \propto J_\perp$. \\

For the intra-leg potential, there is no influence between the two holes. Therefore, the potential is simply the sum of two copies of the single hole case: $V_{\bsigma,\parallel}(x_1,x_2) = V_{\bsigma,\parallel}(x_1) + V_{\bsigma,\parallel}(x_2)$, with [Ref. \cite{Nielsen2023_3}, Eq. (13)]
\begin{align}
V_{\bsigma,\parallel}(x_i) = J_\parallel \left\{\begin{matrix}
\sigma_{(i,-1)}\sigma_{(i,1)} - \sigma_{(i,x_i)}\sigma_{(i,x_i+1)}, & x_i > 0, \\
\sigma_{(i,-1)}\sigma_{(i,1)} - \sigma_{(i,x_i)}\sigma_{(i,x_i-1)}, & x_i < 0, \\
\end{matrix}\right. 
\end{align}
for $i = 1,2$. From here, it is already clear that the potential is not translationally invariant. \\

The more interesting contribution comes from the trans-leg part, $V_{\bsigma,\perp}$. The key observation here, is that if the holes have the same $x$-coordinate, sitting as nearest neighbors across the ladder, any changes in the spin couplings across the ladder has been repaired. This means that $V_{\bsigma,\perp}(x,x) = 0$. It also means that the trans-leg potential \emph{only} depends on the spins between the holes. More explicitly, when $x_1 \neq x_2$, we can first consider $x_1 < x_2$. Here, I break it up into three cases. First, for $x_2 > x_1 \geq 0$
\begin{align}
V_{\bsigma,\perp}(x_1,x_2) = J_\perp (1 - \delta_{x_2,x_1+1}) \sum_{j = x_1+1}^{x_2-1} \sigma_{(1,j)}[\sigma_{(2,j+1)} - \sigma_{(2,j)}] -J_\perp \sigma_{(1,x_2)}\sigma_{(2,x_2)}.
\end{align}
The Kronecker $\delta$ term is there to prevent a contribution from the sum in the case that $x_2 = x_1+1$. Second, for $x_2 > 0 > x_1$
\begin{align}
V_{\bsigma,\perp}(x_1,x_2) =\, &J_\perp (1-\delta_{x_2,+1}) \sum_{j = 1}^{x_2-1} \sigma_{(1,j)}[\sigma_{(2,j+1)} - \sigma_{(2,j)}] - J_\perp\sigma_{(1,x_2)}\sigma_{(2,x_2)} \nn \\
+ &J_\perp (1-\delta_{x_1,-1}) \!\!\!\sum_{j = x_1+1}^{-1} \!\!\!\sigma_{(2,j)}[\sigma_{(1,j-1)} - \sigma_{(1,j)}] - J_\perp\sigma_{(1,x_1)}\sigma_{(2,x_1)} + J_\perp \sigma_{(1,-1)}\sigma_{(2,+1)}.
\end{align}
Third, for $0 \geq x_2 > x_1$
\begin{align}
V_{\bsigma,\perp}(x_1,x_2) = J_\perp (1 - \delta_{x_2,x_1+1}) \sum_{j = x_1+1}^{x_2-1} \sigma_{(2,j)}[\sigma_{(1,j-1)} - \sigma_{(1,j)}] - J_\perp \sigma_{(1,x_1)}\sigma_{(2,x_1)}.
\end{align}
For $x_1 > x_2$, we simply need to swap: $x_1\leftrightarrow x_2$ and $\sigma_{(1,j)}\leftrightarrow \sigma_{(2,j)}$. \\

Finally, let us analyze the potential statistically. Since the spin lattice is assumed to be at infinite temperature, all spin correlation functions vanish: $\braket{\sigma_\bi \sigma_\bj} = 0$ for $\bi \neq \bj$. Therefore, the mean value of the induced interaction vanishes as well: $\braket{V_\bsigma(x_1,x_2)} = \braket{V_{\bsigma,\parallel}(x_1,x_2)} + \braket{V_{\bsigma,\perp}(x_1,x_2)} = 0$. Moreover, the lack of correlations between sites mean that the variance only gets nonzero contributions from onsite terms: $\braket{(\sigma_\bi \sigma_\bj)^2} = \braket{\sigma_\bi^2 \sigma_\bj^2} = 1/16$. As a result, the variance of the intra-leg potential is simply 
\begin{align}
{\rm Var}[V_{\bsigma,\parallel}(x_1,x_2)] = \braket{[V_{\bsigma,\parallel}(x_1,x_2)]^2} = 2 J_\parallel^2 [\braket{(\sigma_{(1,-1)}\sigma_{(1,1)})^2} + \braket{(\sigma_{(1,x_i)}\sigma_{(1,x_i+1)})^2} ] = \frac{J_\parallel^2}{4},
\end{align}
for $x_1, x_2 \neq 0$. If one is at its origin, the variance drops to $\frac{J_\parallel^2}{8}$, and if both are at the origin it vanishes. For the trans-leg potential, we may note that whenever $x_2 \neq x_1$, there is always $2|x_2-x_1| - 1 = 4|\Delta| - 1$ terms of the form $\braket{(\sigma_\bi \sigma_\bj)^2} = 1/16$. Therefore, 
\begin{align}
{\rm Var}[V_{\bsigma,\perp}(x_1,x_2)] = \frac{J_\perp^2}{16}\left[4|\Delta| - 1\right] = \frac{J_\perp^2}{4}\left[|\Delta| - \frac{1}{4}\right].
\end{align}
The total variance is then [see also Eq. (7) of the main text]
\begin{align}
{\rm Var}[V_{\bsigma}(x_1,x_2)] = \frac{J_\perp^2}{4}\left[|\Delta| - \frac{1}{4}\right] + \frac{J_\parallel^2}{4},
\end{align}
whenever both $x_1,x_2$ are nonzero, and when $\Delta = x_1 - x_2 \neq 0$. 

\section{Effective model for $J_\parallel = 0$}
For a vanishing intra-leg spin coupling, $J_\parallel = 0$, and large trans-leg spin coupling $J_\perp \gg t$, it is possible to significantly simplify the description of the dynamics of the two holes. As soon as the two holes move one site apart, this costs an energy of $s J_\perp/4$, where the sign $s=\pm 1$ is fixed by the specific spin realization and thereby randomly chooses $+1$ or $-1$ with equal probability. When $J_\perp \gg t$, we may, thus, assume that the holes never move \emph{more} than one site apart. This leads to the effective one-dimensional model
\begin{align}
\Ham_{\rm eff}(\bsigma) = \sqrt{2}t \sum_{n = -(N_x - 1)}^{N_x - 2} [\hat{p}^\dagger_{n+1}\hat{p}_n + {\rm H.c.}] + \frac{J_\perp}{4}\sum_{\substack{n = -(N_x - 1) \\ n \; {\rm odd}}}^{N_x - 1} s_n(\bsigma) \hat{p}^\dagger_n \hat{p}_n,
\label{eq.effective_model}
\end{align} 
where $s_n(\bsigma)$ again depends on the specific spin realization, $\bsigma$. In this manner, the two-hole dynamics on a two-leg ladder of length $N_x$ is effectively reduced to a 1D chain with length $2N_x - 1$. In this model, all even sites correspond to a hole pair sitting on the same rung in the original lattice $\ket{\psi^{\rm eff}_n} = \hat{p}_n^\dagger\ket{0} = \ket{\psi_{n/2,n/2}}$. All odd sites correspond to the symmetric superposition of shifting the holes with respect to each other by one site: $\ket{\psi^{\rm eff}_n} = \hat{p}_n^\dagger\ket{0} = 2^{-1/2}[\ket{\psi_{(n-1)/2 + 1,(n-1)/2}} + \ket{\psi_{(n-1)/2, (n-1)/2 + 1}}]$. Here, I let $\ket{\psi_{n_1,n_2}}$ denote the wave function for holes at sites $n_1,n_2$ in leg $1$ and $2$, respectively, in the original two-leg ladder. The symmetric configuration at the odd sites, hereby, leads to the enhanced hopping amplitude $\sqrt{2}t$. In this manner, I can compute the rms of the center-of-mass at infinite temperature as
\begin{align}
(X_{\rm rms}^{\rm eff})^2 &= \frac{1}{N_S}\sum_{\bsigma\in S} \bra{\psi^{\rm eff}_0}e^{+i\Ham_{\rm eff(\bsigma)}\tau}\hat{X}^2e^{-i\Ham_{\rm eff}(\bsigma)\tau}\ket{\psi^{\rm eff}_0} = \frac{1}{N_S}\sum_{\bsigma\in S} \sum_{n}|\bra{\psi^{\rm eff}_n}e^{-i\Ham_{\rm eff}(\bsigma)\tau}\ket{\psi^{\rm eff}_0}|^2 \bra{\psi^{\rm eff}_n}\hat{X}^2\ket{\psi^{\rm eff}_n} \nn \\
&= \frac{1}{N_S}\sum_{\bsigma\in S} \sum_{n}|\bra{\psi^{\rm eff}_n}e^{-i\Ham_{\rm eff}(\bsigma)\tau}\ket{\psi^{\rm eff}_0}|^2 \left[\frac{n}{2}\right]^2,
\end{align}
In this expression, the spin configurations are sampled by randomly choosing the local signs $s_n$. I also use that the center-of-mass distance of the pair to their original rung is simply $n / 2$. Using this effective model, I can significantly enhance the system size to a few thousand lattice sites. This enables me to investigate dynamics up to timescales of at least $\tau = 10^4 / t$ when $J_\perp \gg t$, and shows that the subdiffusive behavior discovered in the full model is robust on long timescales. The effective model is also seen, in the main text, to perfectly match the full root-mean-square dynamics of the full model for large enough $J_\perp / t$.

\section{Finite-size effects for $J_\parallel = 0$}
\begin{figure}[t!]
\begin{center}
\includegraphics[width=1.0\columnwidth]{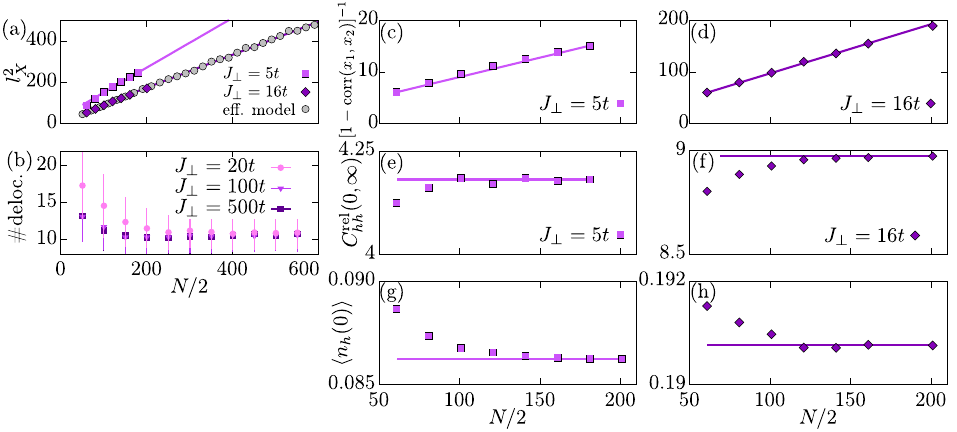}
\end{center}\vspace{-0.5cm}
\caption{(a) Steady state mean-square-distance $l_X^2$ for $J_\parallel = 0$ vs system length, $N/2$, for the full [at specified $J_\perp$] and effective [$J_\perp = 1000t$] models showing $\sim N$ scaling. (b) Number of delocalized states [an IPR scaling as $N$] that has an overlap with the initial state of order $1/N$ in the effective model vs system length. Errorbars denote the standard deviation estimated from $400$ samples. (c),(d) Linear scaling of the inverse correlation coefficient relative to $1$. (e),(f) Saturation of the relative correlator at $\Delta = 0$ vs system length. (g),(h) Saturation of the marginal distribution at the original site of the hole-pair, $x = 0$.}
\label{fig.rms_scaling_and_deloc_states} 
\vspace{-0.25cm}
\end{figure}

Figure \ref{fig.rms_scaling_and_deloc_states}(a) shows the steady state value of the mean-square distance of the center-of-mass, $l_X^2$, vs system length, both for the full and effective model. This shows a clear $\mathcal{O}(N)$ scaling. In the main text, we saw that the pair delocalizes subdiffusively. This means that it must couple to delocalized states. The mean-square distance can formally be calculated using the eigenstates for a given spin realization $\{\ket{E(\bsigma)}\}$ as 
\begin{align}
X_{\rm rms}^2 &= 2^{-N}\sum_{\bsigma} \bra{\Psi_\bsigma(0)}e^{+i\Ham\tau} \hat{X}^2 e^{-i\Ham\tau}\ket{\Psi_\bsigma(0)} \nn \\
&= 2^{-N}\sum_{\bsigma} \sum_{E(\bsigma),E'(\bsigma)} e^{i(E(\bsigma) - E'(\bsigma))\tau} \braket{\Psi_\bsigma(0)|E(\bsigma)}\bra{E(\bsigma)} \hat{X}^2 \ket{E'(\bsigma)}\braket{E'(\bsigma)|\Psi_\bsigma(0)}\to \nn \\
l_X^2 &= 2^{-N}\sum_{\bsigma} \sum_{E(\bsigma)} |\braket{\Psi_\bsigma(0)|E(\bsigma)}|^2\bra{E(\bsigma)} \hat{X}^2 \ket{E(\bsigma)}
\end{align}
Now, the point is that this sum is completely determined by the delocalized states that the initial state has an overlap with, because they will have a mean-square distance of order $N^2$. Hence, we may write
\begin{align}
l_X^2 &= 2^{-N}\sum_{\bsigma} \sum_{E_{\rm del.}(\bsigma)} |\braket{\Psi_\bsigma(0)|E_{\rm del.}(\bsigma)}|^2\bra{E_{\rm del.}(\bsigma)} \hat{X}^2 \ket{E_{\rm del.}(\bsigma)} \nn \\
&= 2^{-N}\sum_{\bsigma} \sum_{E_{\rm del.}(\bsigma)} \frac{C_1(E_{\rm del.}(\bsigma))}{N} [N^2 C_2(E_{\rm del.}(\bsigma))] \nn \\
&= 2^{-N} N \sum_{\bsigma} \sum_{E_{\rm del.}(\bsigma)} C_1(E_{\rm del.}(\bsigma)) \times C_2(E_{\rm del.}(\bsigma)),
\end{align}
where $C_1,C_2$ are coefficients of order $1$. Here, I use that since the states are delocalized $|\braket{\Psi_\bsigma(0)|E_{\rm del.}(\bsigma)}|^2 \propto 1 / N$, and $\bra{E_{\rm del.}(\bsigma)} \hat{X}^2 \ket{E_{\rm del.}(\bsigma)} \propto N^2$. Now, the point is that the effective model reveals that when we sample over the spin realizations, there is only a \emph{constant} number of these delocalized states, i.e. this number does not increase with $N$. This is illustrated in Fig. \ref{fig.rms_scaling_and_deloc_states}(b), showing the number of states with an inverse participation ratio (IPR) $1/\sum_{n}|\braket{\psi^{eff}_n|E}|^4 \sim N$ that has an overlap of order $1/N$ with the initial state in the effective model. This number comes out to around $10$ for large enough $N$ and is independent of $J_\perp / t$ for $J_\perp / t \gg 1$. So, because there is a constant number of these terms for all the spin realizations that are sampled, we get
\begin{align}
l_X^2 &= 2^{-N} N \sum_{\bsigma} \sum_{E_{\rm del.}(\bsigma)} C_1(E_{\rm del.}(\bsigma)) \times C_2(E_{\rm del.}(\bsigma)) \nn \\
&= N \times \frac{1}{N_S} \sum_{\bsigma \in S} \sum_{E_{\rm del.}(\bsigma)} C_1(E_{\rm del.}(\bsigma)) \times C_2(E_{\rm del.}(\bsigma)) \propto N
\end{align}
where I use that $\frac{1}{N_S} \sum_{\bsigma \in S} \sum_{E_{\rm del.}(\bsigma)} C_1(E_{\rm del.}(\bsigma)) \times C_2(E_{\rm del.}(\bsigma))$ is of order $1$, because all $N_S$ terms are of order $1$. This explains why $l_X^2$ is proportional to $N$. It follows directly that the correlation coefficient in the steady state behaves as
\begin{align}
{\rm corr}(x_1,x_2) = \frac{l_X^2 - l_\Delta^2}{l_X^2 + l_\Delta^2} \simeq 1 - \frac{l_\Delta^2}{l_X^2} = 1 - \frac{\alpha}{N},
\end{align}
where $\alpha$ is order $1$. This is essentially because $l_\Delta$ approaches a constant for $N\gg 1$. I have checked this to be true numerically as well. Rephrased, this shows that the correlation coefficient obeys the scaling
\begin{align}
\frac{1}{1 - {\rm corr}(x_1,x_2)} \propto N, 
\end{align}
which is also confirmed numerically in Figs. \ref{fig.rms_scaling_and_deloc_states}(c) and \ref{fig.rms_scaling_and_deloc_states}(d). On the contrary, however, Figs. \ref{fig.rms_scaling_and_deloc_states}(e) and \ref{fig.rms_scaling_and_deloc_states}(f) show that the relative hole-hole correlator 
\begin{align}
C_{hh}^{\rm rel}(\Delta, \tau) = \frac{\sum_X C_{hh}(X + \Delta, X - \Delta; \tau)}{\sum_X \braket{\hat{n}_h(X + \Delta)}\braket{\hat{n}_h(X - \Delta)}} 
\end{align}
saturates at a finite value vs system size [for $\Delta = 0$]. This is a bit counter-intuitive, as one would expect $\braket{\hat{n}_h} \sim 1/\sqrt{N}$ for $\sqrt{N}$ sites (remember that the holes delocalize over $l_X \sim \sqrt{N}$ sites), such that the the disconnected correlator in the denominator should scale as $(1/\sqrt{N})^2 \times \sqrt{N} = 1/\sqrt{N}$. This would mean that $C_{hh}^{\rm rel} \propto \sqrt{N}$, which contradicts the numerical evidence. However, this can be understood by checking how the marginal distribution looks around their original site. Hence, I plot $\braket{n_h(x = 0)}$ in Figs. \ref{fig.rms_scaling_and_deloc_states}(g) and \ref{fig.rms_scaling_and_deloc_states}(h). This is clearly seen to saturate for increasing $N$.

\begin{figure}[t!]
\begin{center}
\includegraphics[width=0.8\columnwidth]{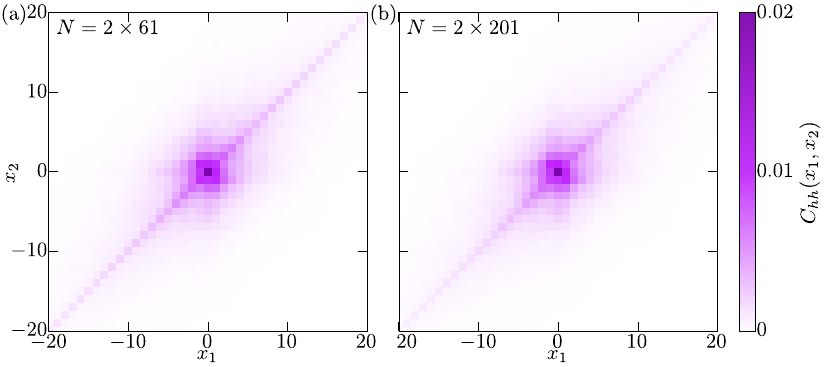}
\end{center}\vspace{-0.5cm}
\caption{Hole-hole correlator for $N = 2\times 61$ (a) and $N = 2\times 201$ (b) at long times [$\tau = 2000/t$] for $J_\perp = 5t$ and $J_\parallel = 0$. Their is a nonvanishing localized core around $(x_1,x_2) = 0$.}
\label{fig.hole_hole_correlator} 
\vspace{-0.25cm}
\end{figure} 

The underlying reason for this saturation is that the joint probability distribution of the holes, i.e. the full hole-hole correlator $C_{hh}(x_1,x_2)$ retains a localized \emph{core} around $(x_1, x_2) = (0,0)$, while there is also a delocalized part around the diagonal, $x_1 = x_2$. This is shown explicitly in Figs. \ref{fig.hole_hole_correlator}(a) and \ref{fig.hole_hole_correlator}(b), where one should notice that while the diagonal probability around $x_1 = x_2$ decreases for the larger value of $N$, the core around $(x_1, x_2) = (0,0)$ is essentially unchanged. 

\section{Soft core potential}
The proposal for the experimental setup is based on Refs. \cite{Zeiher2016,Weckesser2024} using ${}^{87}$Rb. The $\sigma^+$ transition between the $\ket{F = 2, m_F = 2}$ state in the $5S_{1/2}$ manifold and the $\ket{J = 3/2, m_J = 3/2}$ Rydberg state in the $30P_{3/2}$ manifold is \emph{stroboscopically} driven with Rabi frequency $\Omega$, detuning $\delta$, and duty circle $D = 1/600$. The assumed lattice spacing is $a = 752{\rm nm}$, and I set $\Omega = 2\pi \times 20 {\rm MHz}$. I analyze the dynamics for three distinct experimentally feasible values of the detuning, $\delta = 2\pi \times 60 {\rm MHz}, 2\pi \times  80 {\rm MHz}, 2\pi \times  100 {\rm MHz}$. The pulsed scheme dramatically improves the lifetime of the Rydberg-dressed state, while the effective Hamiltonian describing the dynamics remains unchanged \cite{Weckesser2024}. 

\begin{figure}[t!]
\begin{center}
\includegraphics[width=1.0\columnwidth]{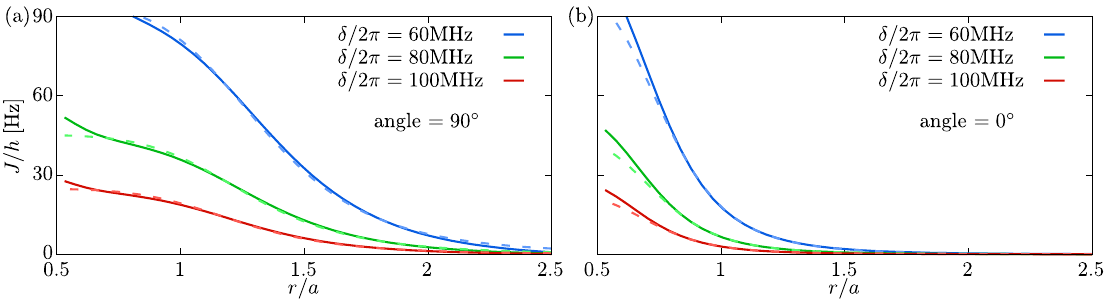}
\end{center}\vspace{-0.5cm}
\caption{(a) Interaction potential in units of Hz as a function of distance in units of the lattice spacing perpendicular to the direction of the applied magnetic field. This is shown for indicated values of the detuning, $\delta$. The bigger the detuning, the smaller the overall energy scale, and the shorter the range of the interaction. (b) Same as in (a), but along the direction of the magnetic field. In the latter case, the potential falls off much more quickly. Dashed lines show approximate potentials by fitting to $J(r) = J_0 / (1 + (r / r_c)^6)$. }
\label{fig.dressed_rydberg_interactions} 
\vspace{-0.25cm}
\end{figure} 

As a first step to calculate the interaction of the dressed atoms, one first needs the pair interaction of two pure Rydberg states $\ket{rr}$. This is performed using the open source code documented in Ref. \cite{Weber2017}. As a next step, the Rabi coupling to these Rydberg states results in a dressing of the spin-$\ket{\uparrow}$ states, $\ket{\tilde{\uparrow}}$. As a result of the dressing, there is an effective interaction potential in the $\ket{\tilde{\uparrow}\tilde{\uparrow}}$ sector. This calculation was carrried through by Pascal Weckesser and follows the recipe given in the Supplementary Material of Ref. \cite{Zeiher2016}. The resulting Hamiltonian [Eq. (14) of the main text]
\begin{equation}
\Ham_J = \frac{1}{2}\sum_{\bi \neq \bj} J_{\bi-\bj} \hat{n}_{\uparrow\bi}\hat{n}_{\uparrow\bj}
\label{eq.H_J_rydberg}
\end{equation}
describes spin-specific density-density interactions. The behavior of this potential depends very much on the orientation of the electron orbital in the addressed $30P_{3/2}$ Rydberg state. This may be manipulated by applying a magnetic field, whose direction defines the quantization axis, and thereby, the orbital orientation. Indeed, the dependency of the potential, $J_{\bi-\bj}$, on the interparticle distance, $r = |\bi - \bj|$, is significantly different in the plane of the orbital (perpendicular to the magnetic field) [Fig. \ref{fig.dressed_rydberg_interactions}(a)], and along the applied magnetic field [Fig. \ref{fig.dressed_rydberg_interactions}(b)]. In the latter case, it falls off much more quickly. By putting the magnetic field \emph{along} the ladder, one may then suppress the interaction in this direction, while the interaction along the rungs is unchanged. 
\begin{figure}[t!]
\begin{center}
\includegraphics[width=0.83\columnwidth]{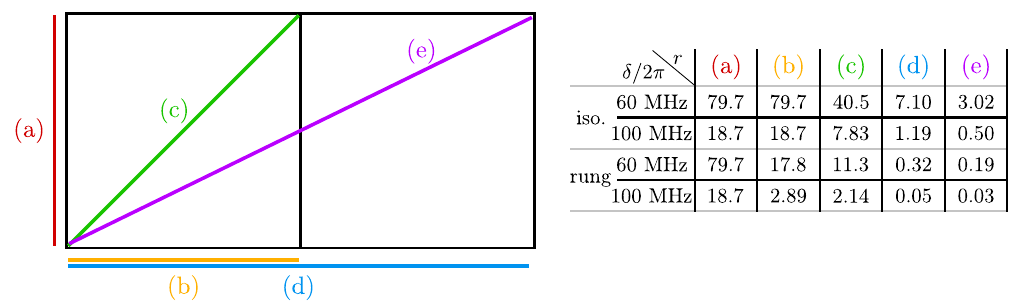}
\end{center}\vspace{-0.5cm}
\caption{The interaction potential is calculated (right) for the shown distances in the lattice (left). Beyond a distance of $\sqrt{5}$ lattice spacings (e), the potential is reduced by $3$ orders of magnitude and neglected. The top two rows in the table refer to the isotropic case, where the magnetic field is perpendicular to the ladder. The lower two rows refer to the rung case, where the magnetic field is along the ladder, and the interaction is suppressed away from the rung.}
\label{fig.calculated_interactions} 
\vspace{-0.25cm}
\end{figure} 
In Fig. \ref{fig.calculated_interactions}, I show for which distances the interaction potential is calculated. In the isotropic case, all distance vectors are at a $90^{\circ}$ angle with the magnetic field and these all follow from the calculation of the potential in Fig. \ref{fig.dressed_rydberg_interactions}(a). In the rung case, the magnetic field points along the ladder. The case (a) in Fig. \ref{fig.calculated_interactions} is hereby still at a $90^{\circ}$ angle and gives the same result as in the isotropic case at that distance. However, cases (b) and (d) must be extracted from the $0^{\circ}$ calculation, while cases (c) and (e) are extracted from calculations at $45^{\circ}$ and $26.6^{\circ}$ with respect to the magnetic field. Importantly, by increasing the detuning from $\delta = 60$ MHz to $\delta = 100$ MHz, the relative size of the interactions at distances (b) and (c) are decreased from $17.8/79.7 \simeq 0.22$, $11.3/79.7 \simeq 0.14$ to $2.89/18.7 \simeq 0.15$, $2.14/79.7 \simeq 0.11$, respectively.
\begin{figure}[t!]
\begin{center}
\includegraphics[width=1.0\columnwidth]{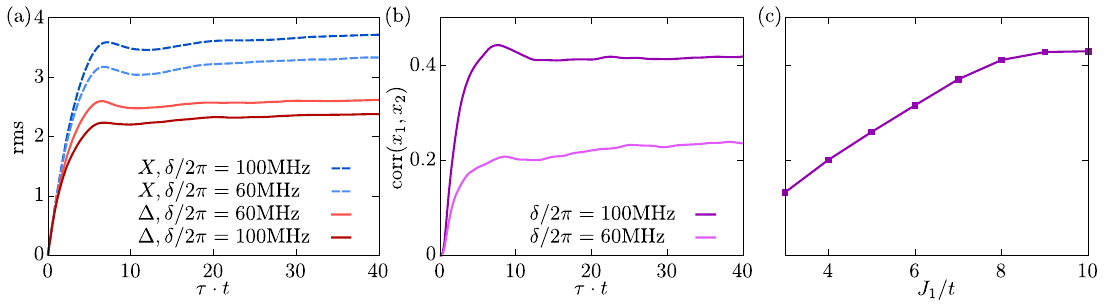}
\end{center}\vspace{-0.5cm}
\caption{(a) The rms dynamics of the center-of-mass, $X$, and relative, $\Delta$, coordinates for indicated detunings, $\delta$, and a magnetic field along the ladder (dominant rung coupling). Note that for lower detunings, the signal worsens both because the rms of the center-of-mass position decreases and because the relative distance increases. (b) The associated dynamics of the correlation coefficient showing much better visibility at larger detuning. (c) Asymptotic correlation coefficient vs $J_1 = J_{{\bf e}_y}$. }
\label{fig.rydberg_dynamics_comparison} 
\vspace{-0.25cm}
\end{figure} 
Although this is a minor change, this actually makes a big difference for the visibility of the results at the chosen system size of $21\times2$. This is shown explicitly in Fig. \ref{fig.rydberg_dynamics_comparison}. Figure \ref{fig.rydberg_dynamics_comparison}(a) shows the rms distance of the center-of-mass and relative coordinates in the two cases. For $\delta / 2\pi = 100$ MHz, there is a much better separation of these two length scales. Indeed, Fig. \ref{fig.rydberg_dynamics_comparison}(b) shows that the correlation coefficient is roughly doubled by going from $\delta / 2\pi = 60$ MHz to $\delta / 2\pi = 100$ MHz. In these calculations, I assume that the strongest spin coupling (along the rung) is $J_1 = J_{{\bf e}_y} = 8t$, as in the main text. This is achievable by adjusting the potential depth of the optical lattice, which is very much within experimental capabilities \cite{Weckesser2024}. Finally, in Fig. \ref{fig.rydberg_dynamics_comparison}(c), I show the asymptotic correlation coefficient for the chosen system size as a function of $J_1 / t$. We see that it levels off around the chosen value of $J_1/t = 8$.

\bibliographystyle{apsrev4-1}
\bibliography{ref_pairing_by_disorder}